\documentclass[a4paper,fleqn,usenatbib]{mnras}

\usepackage[T1]{fontenc}
\usepackage{ae,aecompl}

\usepackage{graphicx}	
\usepackage{amsmath}	

\usepackage{amssymb}	

\title[3D mapping of the ISM X-ray absorption]{3D mapping of the neutral X-ray absorption in the local interstellar medium: The Gaia and XMM-Newton synergy}

\author[Gatuzz et al.]{
Efra\'in~Gatuzz,$^{1,2}$,\thanks{E-mail: 	egatuzzs@eso.org}
S. Rezaei Kh.$^{3}$,
Timothy~R.~Kallman$^{4}$,
Annika Kreikenbohm$^{5,6}$,\newauthor
Mirjam Oertel$^{5}$,
J. Wilms$^{5}$,
and  Javier~A.~Garc\'ia$^{7,5}$ 
\\
$^{1}$ESO, Karl-Schwarzschild-Strasse 2, 85748 Garching bei M\"unchen, Germany\\
$^{2}$Excellence Cluster Universe, Boltzmannstr. 2, D-85748, Garching, Germany\\
$^{3}$Max Plank Institute for Astronomy (MPIA), K\"onigstuhl 17, 69117 Heidelberg, Germany\\
$^{4}$NASA Goddard Space Flight Center, Greenbelt, MD 20771, USA\\
$^{5}$Dr. Karl Remeis-Observatory and Erlangen Centre for Astroparticle Physics, Universit\"at Erlangen-N\"urnberg, Sternwartstr. 7, \\96049 Bamberg, Germany\\
$^{6}$Lehrstuhl f\"ur Astronomie, Universit\"at W\"urzburg, Campus Hubland Nord, Emil-Fischer-Strasse 31, 97074 W\"urzburg, Germany\\
$^{7}$Cahill Center for Astronomy and Astrophysics, California Institute of Technology, Pasadena, CA 91125, USA\\
}

\date{Accepted XXX. Received YYY; in original form ZZZ}

\pubyear{2016}

\begin{document}
\label{firstpage}
\pagerange{\pageref{firstpage}--\pageref{lastpage}}
\maketitle

\begin{abstract}
We present a three-dimensional map of the hydrogen density distribution in the Galactic interstellar medium. The hydrogen equivalent column densities were obtained from the Exploring the X-ray Transient and variable Sky project ({\sc EXTraS}) which provides equivalent $N_{\rm H}$ values from X-ray spectral fits of observations within the  {\it XMM-Newton} Data Release. {\sc EXTraS} include multiple fits for each source, allowing an accurate determination of the equivalent column densities, which depends on the continuum modeling of the spectra. A cross-correlation between the {\sc EXTraS} catalogue and the first {\it Gaia} Data Release  was performed in order to obtain accurate parallax and distance measurements. We use a Bayesian method explained in \citet{rez17} in order to predict the most probable distribution of the density at any arbitrary point, even for lines of sight along which there are no initial observation. The resulting map shows small-scale density structures which can not been modeled by using analytic density profiles. In this paper we present a proof of concept of the kind of science possible with the synergy of these catalogs. However, given the systematic uncertainties connected to the source identification and to the dependence of $N_{\rm H}$ on the spectral model, the present maps should be considered qualitatively at this point.   
\end{abstract}

 
\section{Introduction}\label{sec_int}

The interstellar medium (ISM) includes cold (<10$^{4}$ K), warm (10$^{4}$ -- 10$^{6}$ K) and hot (>10$^{6}$ K) components \citep[see][and references therein]{dra11}. Neutral hydrogen is contained in the warm neutral medium (WNM) and cold neutral medium (CNM).  The CNM component has been mapped using 21~cm emission, on an angular scale of 0.6~degrees together with distance information to be inferred from velocities and  assumed galactic rotation curves \citep{kal05,win16}. Molecules are associated with the CNM and can be detected via their valence electronic transitions \citep{car70} and short wavelength radio line emission \citep{dam01}. The column densities of ions in the WNM are typically probed using absorption by ultraviolet (UV) resonance lines from cosmically abundant elements along the line of sight (l.o.s) to hot stars \citep{jen11}. Such probes have been carried out for a few hundred l.o.s out to distances less than a few kpc. The hot ionized medium (HIM) can be probed using soft X-ray emission \citep{sno97, hen10}, but owing to its  path length through the galactic disc it can be observed only for a fraction of the galaxy.  Absorption by the HIM can be seen in high ionization resonance  lines from ions such as {\rm O}~{\sc vii} and {\rm Ne}~{\sc ix} \citep{gat17}. The warm ionized medium (WIM) can be studied  using H$\alpha$ emission revealing a larger scale height than for the neutral components, together with different discrete structures \citep{haf03}. The cold phase of the ISM, finally, plays an important role in the Galactic evolution acting as a gas reservoir for star formation processes \citep{san08}. This cold component is dominated by Hydrogen, the most abundant chemical element in the universe. Therefore, an study of the {\rm H} spatial distribution along the Milky Way is key to understanding the chemical evolution of our own Galaxy. Moreover, accurate measurements of Galactic {\rm H} are required by a variety of analysis such as 3D hydrodynamic galaxy simulation, X-ray dust scattering and X-ray binaries population.  

Multiple all-sky surveys have been performed in the last decades, showing that most of the {\rm H}~{\sc i} gas resides in a thin disc along the Galactic plane, with the Sun embedded within it \citep{dic90,har97,baj05,kal05}.  Hence, all l.o.s from the earth to a point in the sky show a certain degree of absorption, with hydrogen column densities varying by a factor of 100 or more. Surveys using both radio and H$\alpha$ have limited angular resolution, typically $\sim$0.25 deg$^2$ or greater.  Finer scale spatial structure requires targeted  surveys over smaller patches of the sky, or absorption studies toward point sources. Emission studies (of collisionally excited radiation) are sensitive to density fluctuations via the density squared dependence on emissivity; while absorption analysis is more democratic owing to the linear dependence of opacity on density.

 Although several studies have been performed to model the ISM X-ray absorption using high-resolution spectra \citep{sch02, tak02,jue04,jue06,yao09,lia13,pin10,pin13,gat13a,gat13b,gat14,gat15,gat16,nic16a,gat17,gat18b} a detailed analysis of the ISM using the moderate-resolution provided by the CCD cameras on board of {\it XMM-Newton} has not been performed yet. The main advantage of such study is that moderate resolution spectra can be analyzed faster, in comparison to the detailed analysis required in the presence of multiple absorption lines, while providing a good measure of the equivalent $N({\rm H})$ absorption. Moreover, compared with the angular resolution provided by the 21~cm all-sky surveys ($\sim 36$ arcmin) the {\it XMM-Newton} angular resolution (about 6 arcsec FWHM) allows a study of the small-scale ISM structure. The cross section associated with unit extinction by dust $E(B-V)$=1 is  1.7 $\times 10^{-22}$ cm$^{-2}$. As a practical matter, absorbing columns can be measured most easily for spectra of stars with $E(B-V)$ $\leq$ 0.1. Thus studies of optical-UV extinction can study the interstellar medium with column densities 10$^{18}$ -- 10$^{21}$ cm$^{-2}$. The X-ray photoelectric cross section, on the other hand, is approximately 2 $\times 10^{-22} ({\rm E}/1 {\rm keV})^{-3}$ cm$^{-2}$ per hydrogen atom \citep{bro70}; this value depends weakly on the ionization state or metallicity of the gas. In this sense, the X-rays generally can probe columns and distances larger than the optical measurements. 

 In this paper we present a study of the equivalent $N({\rm H})$ distribution in the Milky Way obtained from {\it XMM-Newton} spectral fitting, using the distance measurements obtained by {\it Gaia}. The angular resolution provided by {\it XMM-Newton} allows an analysis of small-scale structures in the cold ISM.  The outline of the present paper is as follows. In Section~\ref{sec_dat} we describe the sample of Galactic objects selected from the {\it XMM-Newton} spectral-fit database. The method to create a 3D mapping of the ISM absorption is described in Section~\ref{sec_map}. Section~\ref{sec_21cm} shows a comparison with results obtained from 21~cm surveys followed by a discussion of the caveats and limitations in our method in Section~\ref{sec_cav}. Finally, the conclusions are summarized in Section~\ref{sec_con}.
 
 \section{Galactic objects sample}\label{sec_dat}
The Exploring the X-ray Transient and variable Sky project ({\sc EXTraS}\footnote{http://www.extras-fp7.eu/}) provides a catalogue of X-ray EPIC PN/MOS spectral fitting parameters for observations within the {\it XMM-Newton} catalogue. The current version of {\sc EXTraS} includes 137212 observations from the {\it XMM-Newton} Data Release \citep[3XMM-DR6,][]{ros16}.  The following spectral models are included in the catalogue:

\begin{itemize}
\item Simple models:
\begin{enumerate}
\item An absorbed power-law model ({\sc XSPEC}: {\tt tbnew*pow}).
\item An absorbed thermal model ({\sc XSPEC}: {\tt tbnew*apec}).
\item An absorbed black-body model ({\sc XSPEC}: {\tt tbnew*bbody}).
\end{enumerate}

\item Complex models:
\begin{enumerate}
\item An absorbed emission spectrum from collisionally-ionized diffuse gas plus power-law model ({\sc XSPEC}: {\tt tbnew*(apec+tbnew*pow)}).
\item An absorbed double power-law model ({\sc XSPEC}: {\tt tbnew*(pow+tbnew*pow)}).
\item An absorbed black-body plus power-law model ({\sc XSPEC}: {\tt tbnew*(bbody+pow)}).
\end{enumerate}
\end{itemize}

Here {\tt tbnew} is a revised version of the X-ray absorption model of \citet{wil00}. The models are selected to represent the most commonly observed spectral shapes in astronomical sources in a phenomenological way. The complex models defined above are only fitted for observations with $\geq 500$ counts in the total energy band. We have computed EPIC-PN spectra simulations in order to estimate the expected uncertainties in the column densities for sources with 500 counts in the complete energy range (0.5-10~keV). We found that, in the case of the simple models described above, for  a column density of $N({\rm H})=10^{21}$ cm$^{-2}$ we obtain a column density uncertainty of $\sim$25$\%$ (for {\tt powerlaw} and {\tt blackbody} models) and $\sim$30$\%$ (for the {\tt apec} model). For more complex models the $N({\rm H})$ uncertainty can be up to $\sim$40$\%$. The uncertainty becomes larger as the number of counts decrease. For example, for a source with 100 counts we found an uncertainty of $\sim$80$\%$  for a simple model and $>$100$\%$ for a complex one.

When multiple observations are available for the same source, each observation is fitted separately. Each source observation can include either PN data only, PN plus MOS (1 \& 2 detectors) or MOS (1 \& 2 detectors) data only. In cases where multiple detectors are provided, the spectral fitting includes a constant parameter to account for the different effective areas of the instruments. High-resolution spectra, provided by the Reflection Grating Spectrometer (RGS), are not included in the {\sc EXTraS} catalog. Also, in order to account for low counts regime, cash statistics \citep{cas79} is used to fit the data (implemented as {\tt c-stat} in {\sc xspec}). 

 The optical depth ($\tau$) is obtained through the well known relation:

\begin{align}
&&\tau(E) = N({\rm H})\sum_{i,j} A_{i}\sigma_{i,j}(E)\ ,
\label{eq:tau}
\end{align}
where $A_{i}$ is the abundance of the $i$-element with respect to hydrogen, $\sigma_{i,j}(E)$ is the photoabsorption cross-section of the ion (including photoionization) and $N({\rm H})$ is the hydrogen equivalent column density (i.e. the free parameter in the {\tt tbnew} component). Standard abundances for the ISM from \citet{wil00} were used in all models. In summary, the {\sc EXTraS} catalogue provide equivalent $N({\rm H})$ values by fitting the curvature of the X-ray spectrum with the models described above. For each model, the best-fit parameters, the statistical uncertainty and the fit-statistics (i.e. {\tt c-stat} value) are available.

A phenomenological classification scheme, developed by the {\sc EXTraS} team, allows the identification of a given source according to their spectral properties. The scheme is based on the random forests probabilistic method developed by \citet{bre01} which has been successfully applied to classify X-ray sources \citep{lo14,far15}. The source type categories in the catalog include: Seyfert~1, Seyfert~2, BL~Lac, High Mass X-ray Binaries (HMXB), Low Mass X-ray binaries (LMXB), cataclysmic variables, ultraluminous X-ray source (ULX) and stars. In order to perform the classification, a training sample of well know sources is created first in order to define source types in the algorithm. The total number of sources used in the training sample consist of 2911 sources, from which 563 are stars. Then, the random forest method is applied to the complete {\sc EXTraS} catalog in order to classify the remaining sources. A total of 24709 stars are found. For each source a confidence parameter, the probability for a given source classification, is included (i.e. if confidence = 0.98, the probability for the source to be a `star' is 98$\%$). For the training sample confidence is equal to $1.0$. Given this source classification, we have build an initial sample using sources assigned as `star' with a confidence value $> 0.32$ (1$\sigma$). This initial sample consist of 21946 sources. 

When multiple observations were performed for a given source we have selected the observation with the highest number of counts. Then, in order to select the column densities given by the different models used on the spectral fitting, we used $N({\rm H})$ from the model for which {\tt c-stat}/dof is closest to 1.0. Finally, in cases where multiple models have similar {\tt c-stat}/dof values we have selected the simplest model over the complex ones, giving priority to the black-body model (because we are modeling stars).

The largest available catalogue of distances comes from the {\it Gaia} mission \citep{gai16b}. The {\it Gaia} observatory, launched on 2013, will allow the determination of parallaxes and proper motions for > 1 billion sources in the final data release. The first {\it Gaia} Data Release \citet[{\it Gaia} DR1\footnote{This is a proof-of-concept paper that used DR1. We are in the process of analyzing the DR2 data which became available during the publication of this paper.}][]{gai16a} contains parallax measurements for 2.5 million sources. It is important to note that the estimation of distances needs to be done in a proper way in the sense that inverting the parallax is only valid in the absence of noise and therefore the distance calculation should be treated as an inference problem. Therefore, we used the distances and uncertainties inferred by \citet[][hereafter AST16]{ast16b} which  include the required systematic uncertainty of 0.3 mas\footnote{http://www.mpia.de/homes/calj/tgas\_distances/main.html}.

\begin{figure} 
\includegraphics[scale=0.35]{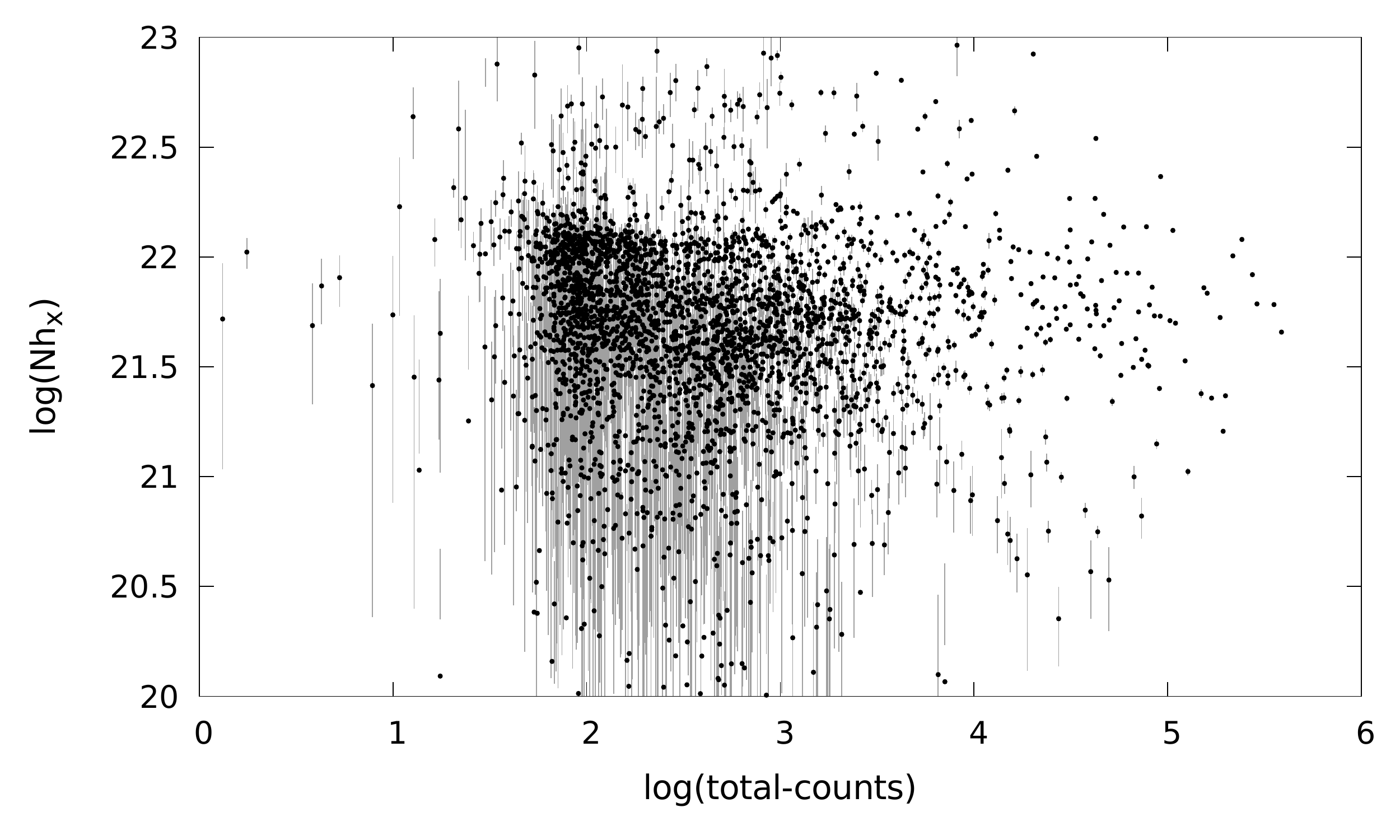}
      \caption{Column densities as function of number of counts after performing the {\sc EXTraS} and AST16 cross-match. Black points indicate the data while grey bars indicate the uncertainties.}\label{fig_error_100}
   \end{figure}

We performed a cross-matching between the {\sc EXTraS} and AST16 catalogues by computing the angular distance between sources. For each source in our sample we assigned the distance obtained from the closest AST16 source. The upper limit in the angular distance is determined from the {\it XMM-Newton} angular resolution (< 12.5 arcsecond). We found a mean angular distance in the cross-matching of 2.1 arcsecond. Figure~\ref{fig_error_100} shows the $N({\rm H})$ densities obtained for those sources for which distance uncertainty is $<20\%$ after the cross-matching process. At this step we noted that (i) numerous sources have large column density errors and (ii) for sources with number of counts $<10^{2}$ we are not able to recover low column densities (<$10^{21}$ cm$^{-2}$). In order to build a final sample, we decide to exclude sources with $\Delta N({\rm H})>50\%$ and number of counts $<10^{2}$. This final sample  consists of 2128 sources and is used in the following analysis.

    \begin{figure*} 
\includegraphics[scale=0.5]{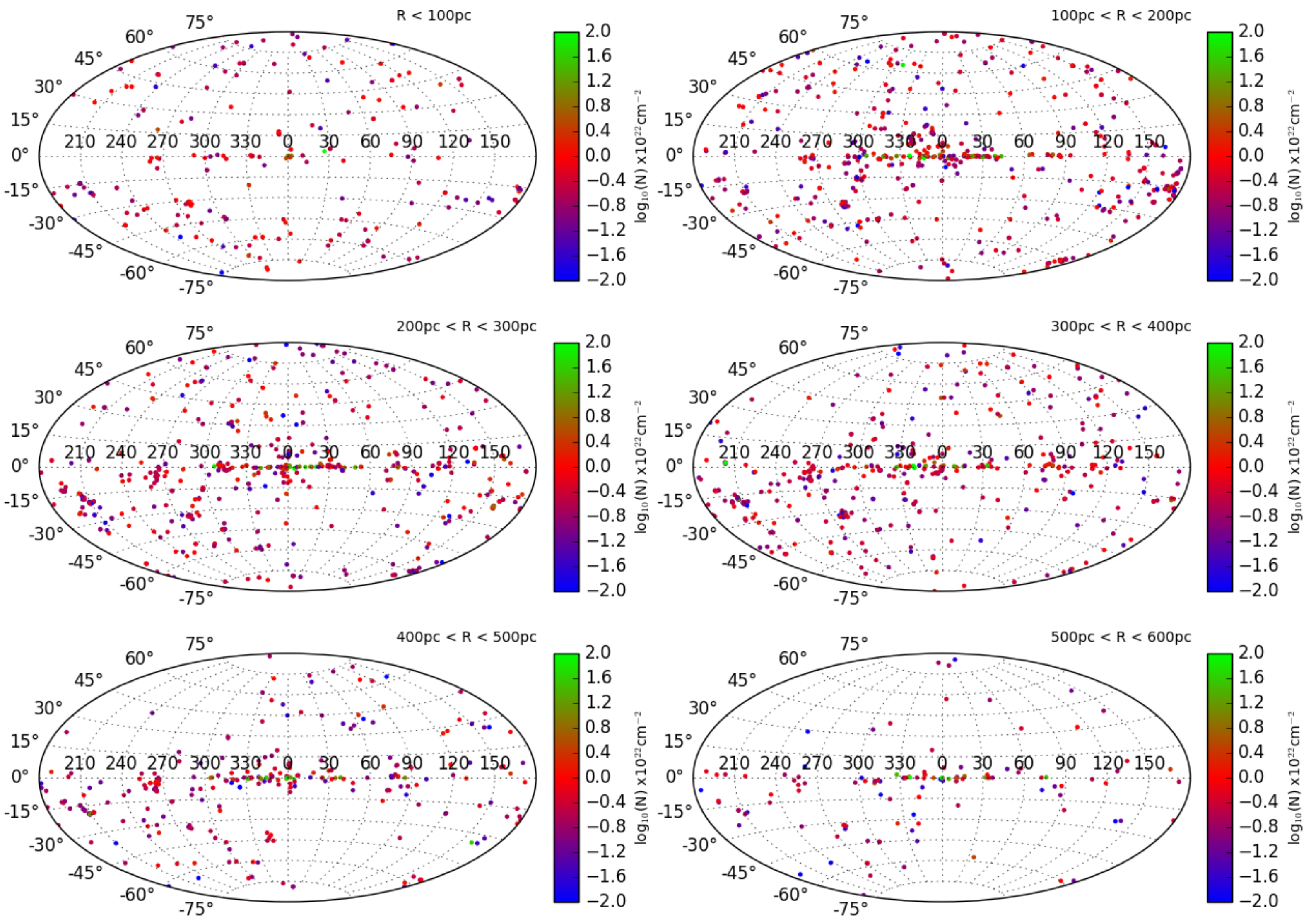}
      \caption{Galactic distribution of the sources in Galactic coordinates.}\label{fig1}
   \end{figure*}

\begin{figure}
\includegraphics[scale=0.35]{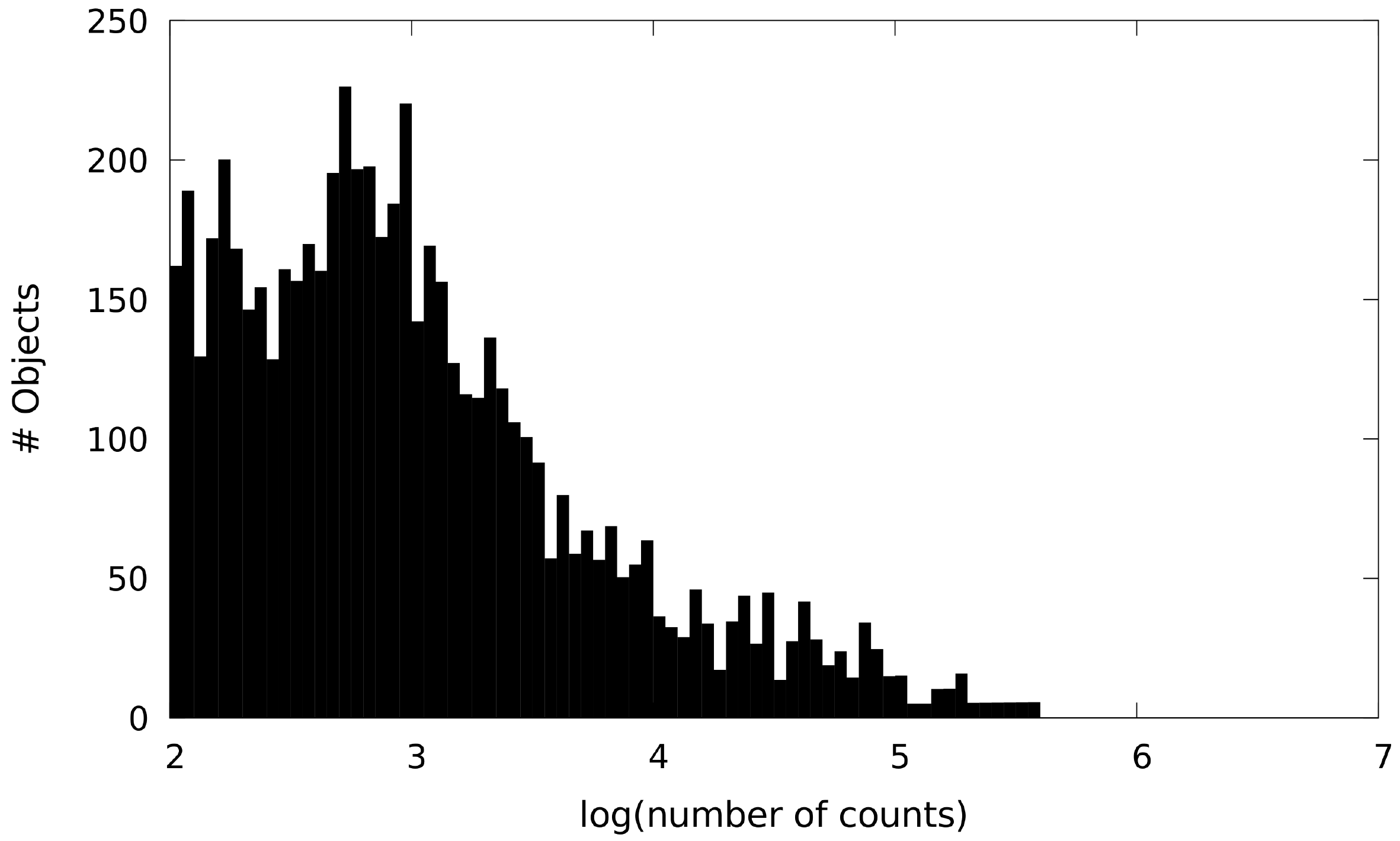}
      \caption{Distribution of the sources according to their number of counts.}\label{fig3}
   \end{figure}

Figure~\ref{fig1} shows the Galactic distribution of the sample in Aitoff projection for different distance ranges. The colors indicate the Hydrogen equivalent column density for each source (obtained from the X-ray fitting procedure described above), in units of 1$\times 10^{22}$ cm$^{-2}$. For illustrative purposes, all sources with equivalent column densities larger than  1$\times 10^{22}$ cm$^{-2}$ show the same color. Although sources are concentrated near the Galactic plane, the sample also includes high latitude sources. Figure~\ref{fig3} shows the distribution of the sources according to the number of counts in the full spectral band (0.5--10 keV).
  
\begin{figure}
\includegraphics[scale=0.35]{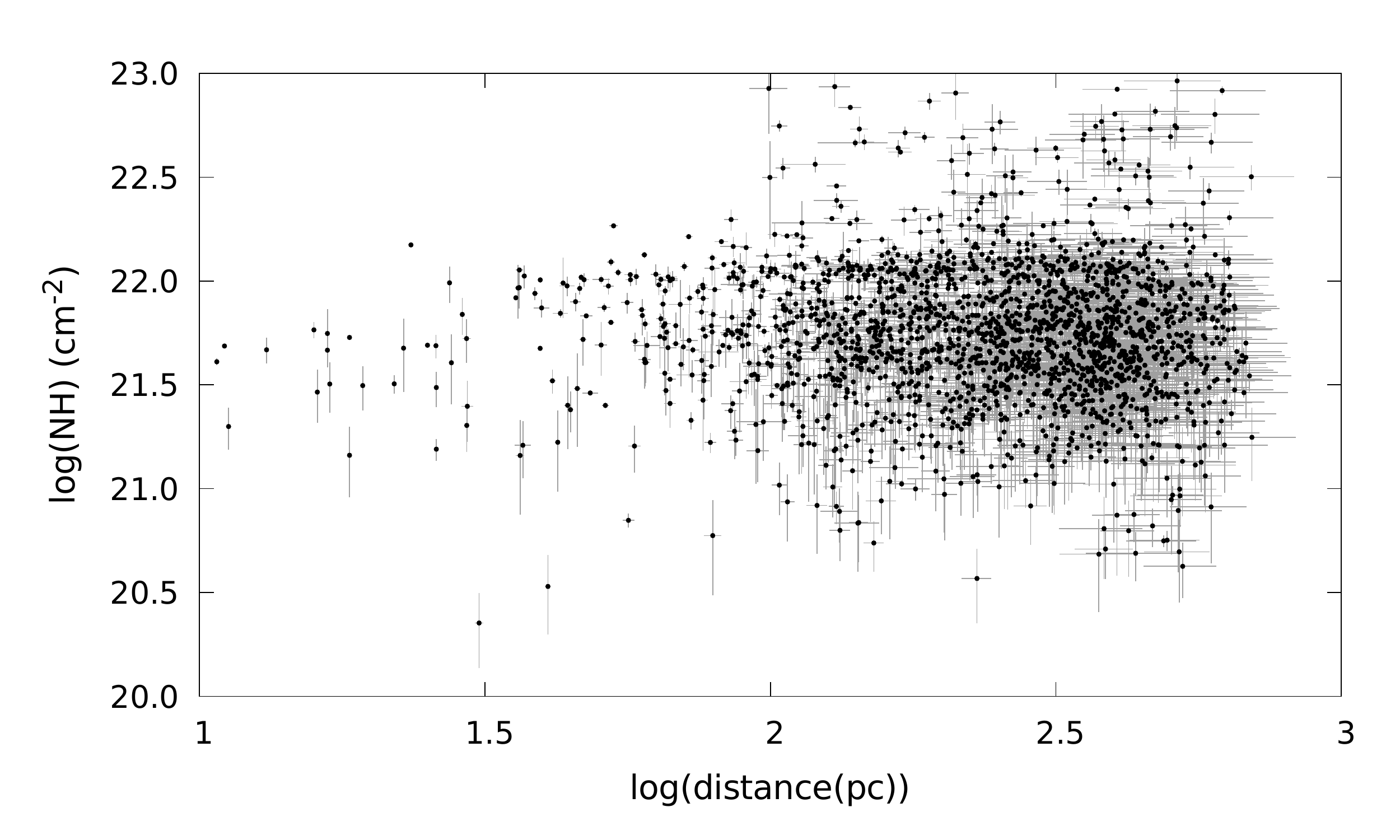}
      \caption{Equivalent $N({\rm H})$ as a function of the distances derived from the cross-matching between {\sc EXTraS} and AST16. Black points indicate the data while grey bars indicate the uncertainties.}\label{fig2}
   \end{figure}

\begin{figure}
\includegraphics[scale=0.35]{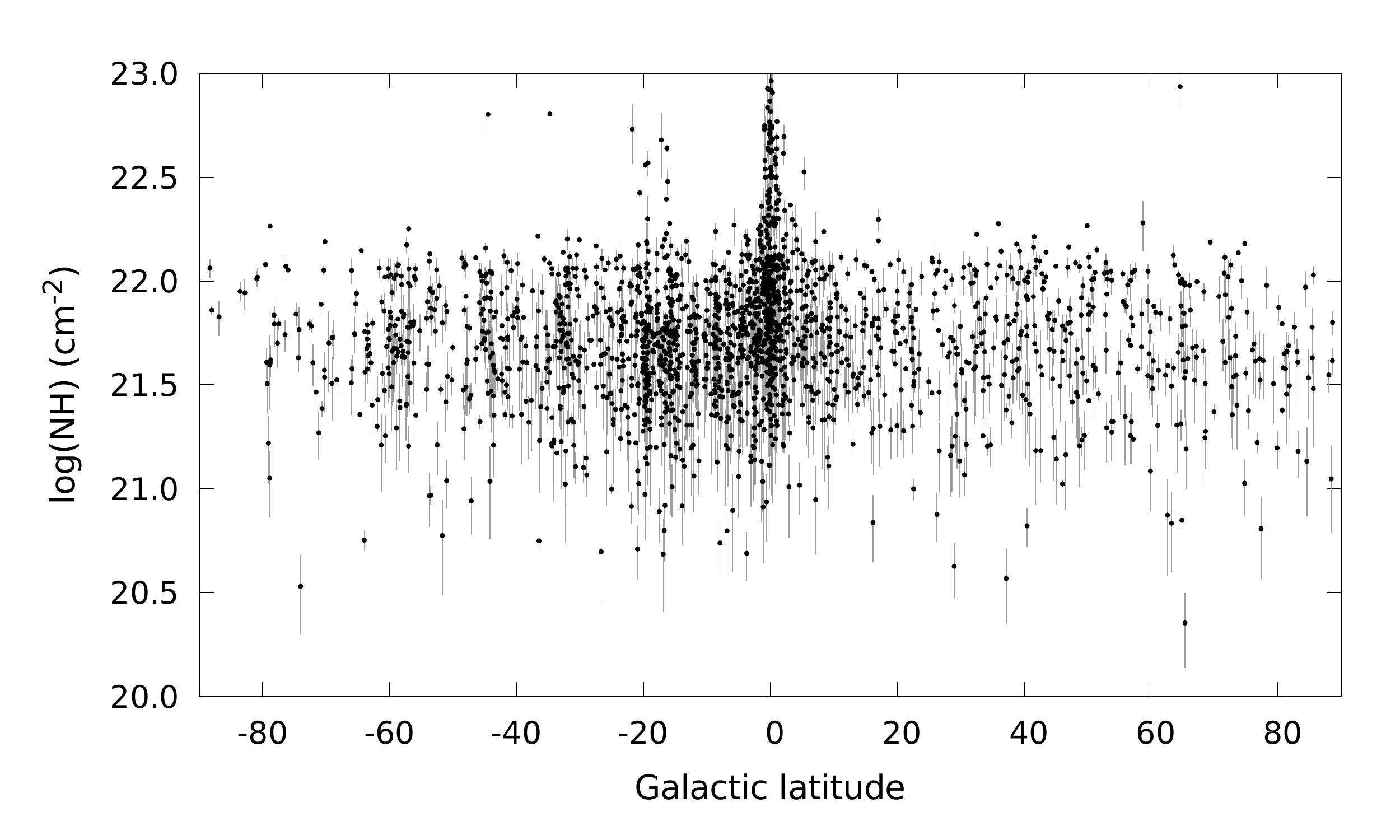}
      \caption{Equivalent $N({\rm H})$ as a function of the Galactic latitude. Black points indicate the data while grey bars indicate the uncertainties.}\label{fig_latitude}
   \end{figure}

Figure~\ref{fig2} shows the equivalent $N({\rm H})$ distribution as a function of the distance for each source in our Galactic sample.  The hydrogen equivalent column density ranges from $10^{20}$ -- $10^{23}$ cm$^{-2}$ for objects with distances between 10~pc and 10~kpc. Figure~\ref{fig_latitude} shows the  $N({\rm H})$ distribution as a function of the Galactic latitude. It is clear from the plot that higher column densities are found near the Galactic plane.

\begin{figure}
\includegraphics[scale=0.35]{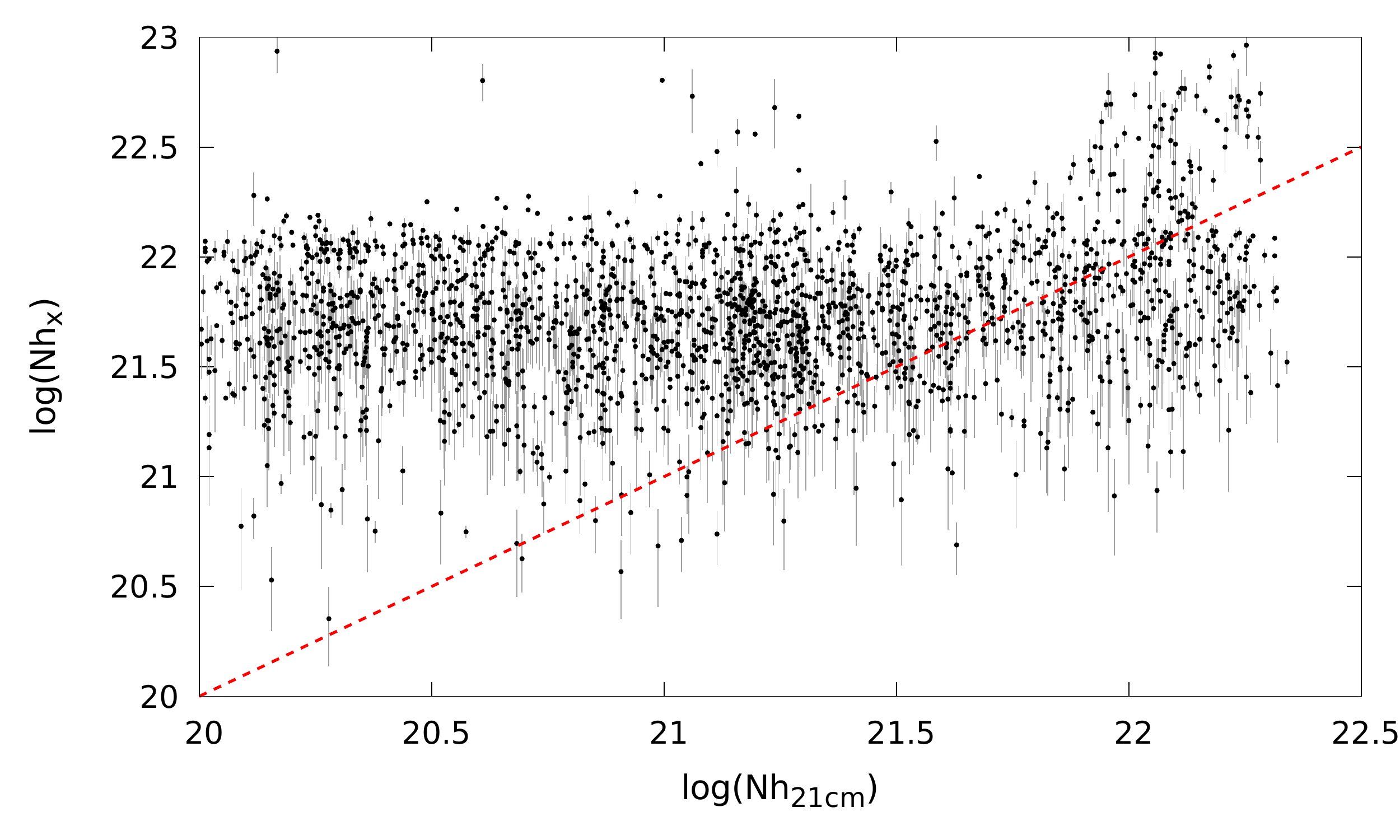}
      \caption{Distribution of neutral equivalent column density derived in X-rays from {\sc EXTraS} vs. \citet{kal05} 21~cm column density. The $\log($NH$_{x})$/$\log($NH$_{21cm})$ $=1$ ratio is plotted with a dashed red line. Black points indicate the data while grey bars indicate the uncertainties.}\label{fig4}
   \end{figure}

\begin{figure}
\includegraphics[scale=0.35]{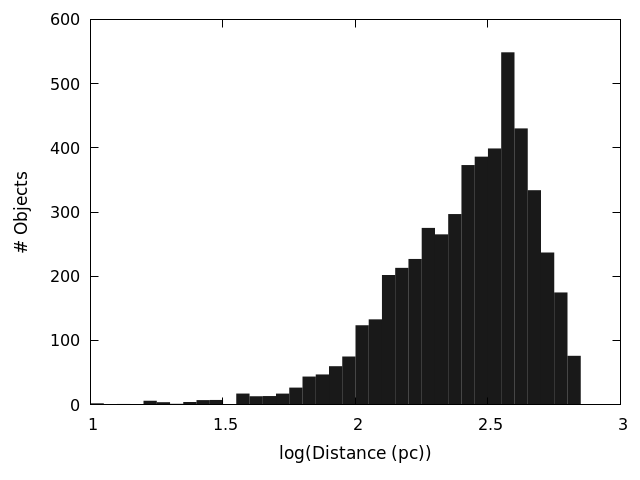}
      \caption{Numbers of sources vs. distance in our final sample. }\label{fig5}
   \end{figure}

Figure~\ref{fig4} shows the distribution of neutral equivalent column density derived in X-rays from {\sc EXTraS} versus the 21~cm column density measurement from \citet{kal05}. It is clear that a significant fraction of the sources show a larger X-ray column than the 21~cm measurements. It is important to note that 21~cm maps provide a measurement over the entire Galactic l.o.s, which can lead to an over-estimation of the column densities for near sources \citep{gat17}.  Figure~\ref{fig5} shows the number of sources versus distance in our sample.  
  
\section{3D mapping of the ISM neutral absorption}\label{sec_map}
Two regimes can be distinguished when studying the {\rm H}~{\sc i} distribution in the Milky Way: large-scale and small-scale structures. Large-scale structure refers to the global distribution of the gas assuming hydrostatic equilibrium conditions. In this scale the density of the gas in the disc can be approximated by an exponential profile \citep{rob03,kal08} although different density profiles have been used to model the spatial distribution such as 
 hyperbolic secant \citep{mar10}  and Gaussian profiles \citep{san84,amo05,bar16}. Using the equivalent $N({\rm H})$ values obtained from the {\sc EXTraS} catalogue the gas distribution can be modeled according to the equation
\begin{align}
&&N_{i}=\int _{\mathrm{observer}}^{\mathrm{source}} n_{i}(r)dr
\end{align}
where $r$ is the distance along the $i$ l.o.s and $n_{i}(r)$ is the density profile. Because our sample does not include sources near the Galactic center or near the outer regions, we cannot constrain the parameters that define the analytic density profiles defined above (e.g. core density $n_{0}$). We use instead a method explained in \citet{rez17} to infer Hydrogen densities from the Hydrogen column density. The method divides each l.o.s towards sources in the 3D space into small 1D pencil beams and uses a Bayesian approach to predict the most probable distribution of the density at any arbitrary point, even for l.o.s along which there are no initial observations. The hyper-parameters of the model are set by the input data and define the type/size of structures which would come out in the density predictions. It is important to note that the concept of cells only exists for setting up the likelihood of the model and the predictions are made for arbitrary points \citep{rez17}. In the following we describe the approach in greater detail.

To begin with, the size of the 1D cells towards each source needs to be set to the typical separation between the input sources, which in our case is $70$ $pc$. Afterwards, there are three hyper-parameters of the Gaussian process which need to be fixed: first is $\lambda$ which is the correlation length. It needs to be a few times the cell sizes so that it can connect the nearby cells. Here we use the correlation length of $500$ $pc$. Second is the variation scale, $\theta$ which sets the maximum fluctuation the model can have to capture the variations. It is calculated based on the variance in the input distribution \citep[see equation 16 in][]{rez17}. We fix this parameter at $2\times10^{-4}$. Finally, since we are using a developed version of the method (Rezaei Kh. et al. submitted), we can set a non-zero mean for the Gaussian process based on the input data, which is $0.007$ $[cm^{-2}pc^{-1}]$. Having set these hyper-parameters we predict the probability distribution function over densities, which is characterized by the mean and the standard deviation, for 50,000 points randomly distributed in the 3D data space. It is important to note that negative values can be obtained through this method which, although they do not have a physical meaning, they provide information about the input data (e.g. very noisy equivalent $N({\rm H})$). For this sample approximately half of the predicted results are negative. These negative values have been set to zero for illustrative purposes.

\begin{figure*}
\includegraphics[scale=0.43]{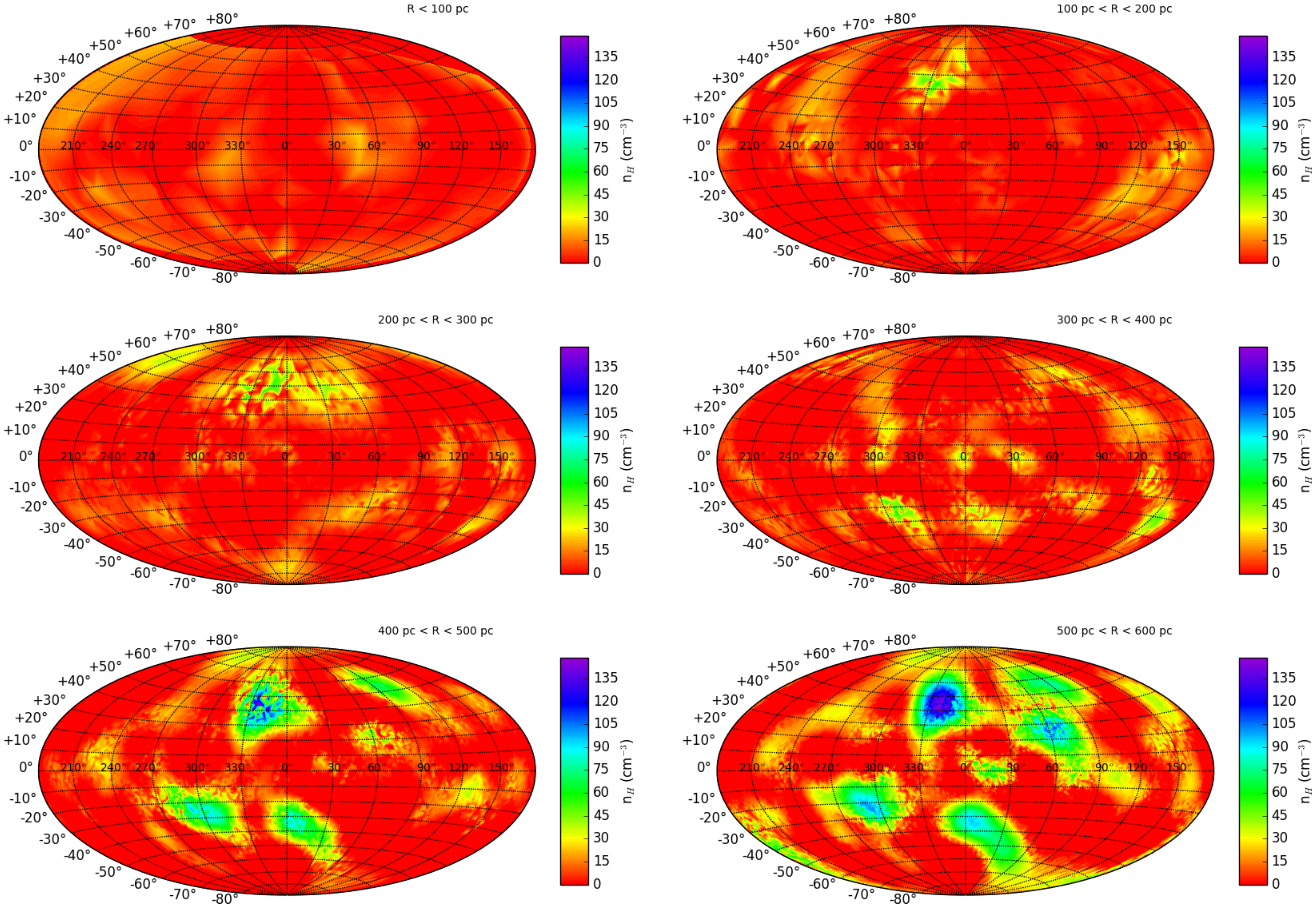} 
      \caption{Full-sky 2D map of the density distribution. The map is an Aitoff projection covering distances from $0$ to $600$ pc.   }\label{fig6}
   \end{figure*}

Figure~\ref{fig6} shows a 2D full-sky map of the density distribution obtained for different distance ranges. The map is in Aitoff projection. We have performed a linear interpolation in order to included those regions for which no points were computed with the method described above. The maximum distance value in our final sample corresponds to $r=600$ pc. In this sense, we are doing an analysis of the very local ISM density distribution, surrounding the Solar System. ${\rm HI}$ intermediate-velocity clouds at high-latitude has been identified by \citet{roh16}, which can explain the high-density region located near (360$^{\circ}$,+40$^{\circ}$) for large distances. However, we noted that density uncertainties increase as we move to the boundaries of the data region (see below).

       \begin{figure}
\includegraphics[scale=0.35]{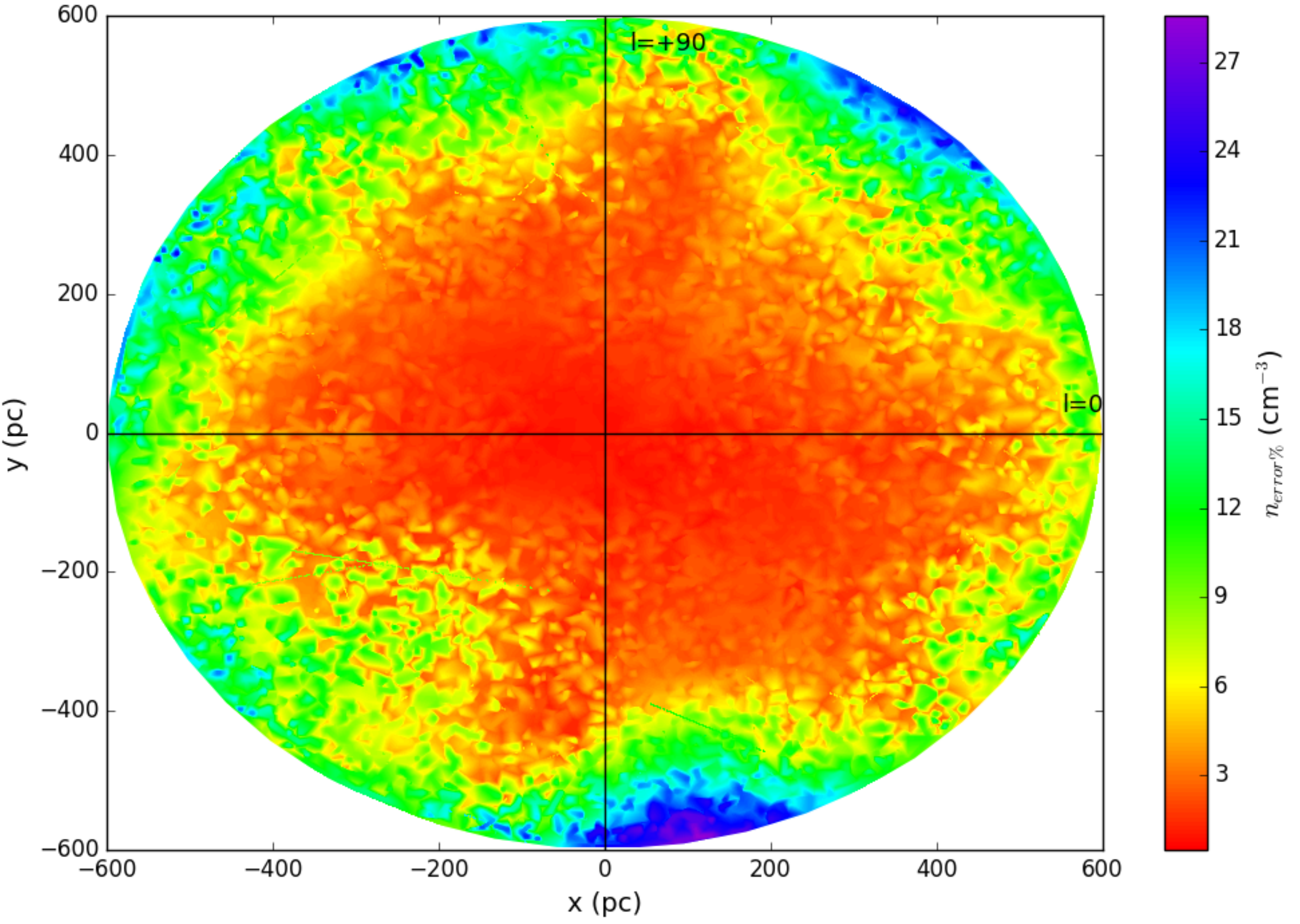} 
      \caption{($x$,$y$) density uncertainties distribution projection map. The Sun is located at coordinates $(x,y)=(0,0)$ and the Galactic center direction at right.}\label{fig_den_error}
   \end{figure}

For plotting purposes we convert the Galactic coordinates to cartesian coordinates ($x$,$y$,$z$) using the distance values obtained from AST16 according to the equations

\begin{align}
&&x=r\cos(l)\cos(b)\\
&&y=r\sin(l)\cos(b)\\
&&z=r\sin(b)
\end{align}
where $r$ is the distance, $l$ is the Galactic longitude and $b$ is the Galactic latitude. We then linearly interpolate the densities to compute a complete density map including regions with no points computed with the method described above. Figure~ \ref{fig_den_error} shows a ($x$,$y$) percentage density uncertainties distribution projection map, normalized to the maximum density value obtained. The uncertainties for the density predictions tend to increase as the predictions get closer to the boundaries of the data volume. This is due to the fact that as we get closer to the boundary, the number of input sources within the correlation volume decreases which makes the predictions more uncertain in these areas. By including more sources from the next {\it Gaia} data releases the uncertainties for the density prediction will decrease.

    \begin{figure*}
\includegraphics[scale=0.36]{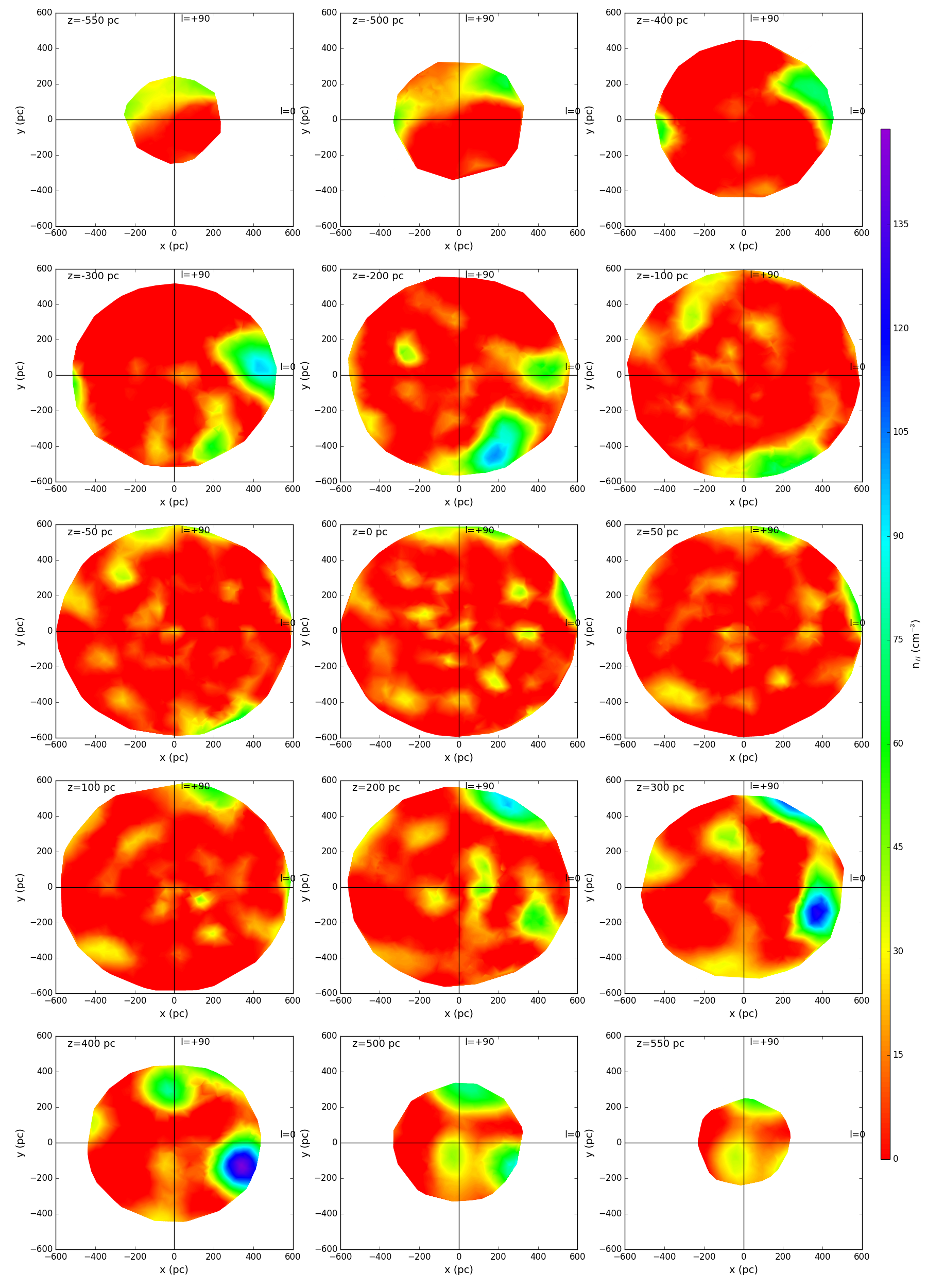} 
      \caption{($x$,$y$) density distribution projection maps for different $z$ values. On each panel, the Sun is located at coordinates $(x,y)=(0,0)$ and the Galactic center direction at right. }\label{fig7}
   \end{figure*}

    \begin{figure*}
\includegraphics[scale=0.3]{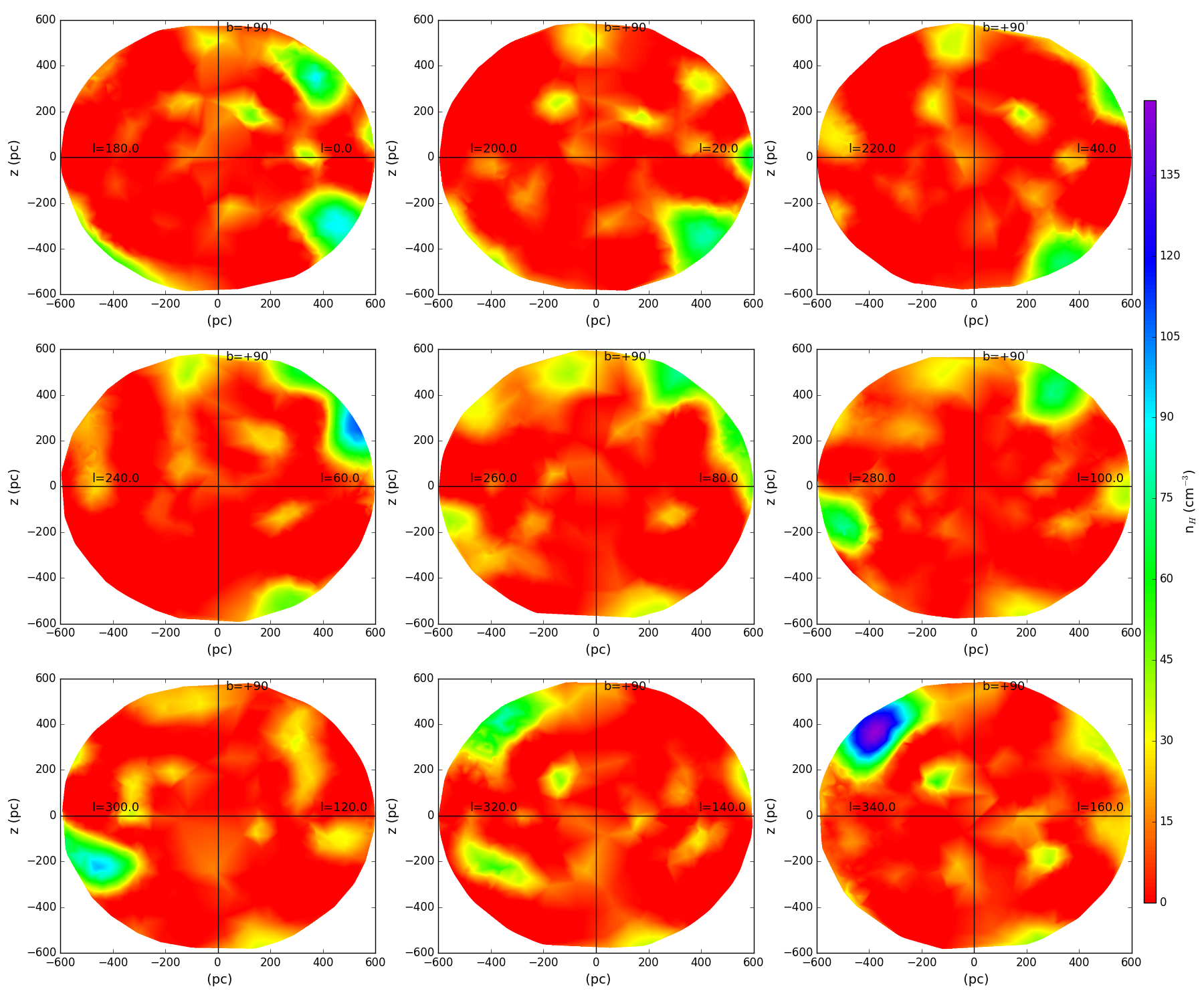} 
      \caption{Vertical slices of the Hydrogen density map for different Galactic longitude values. On each panel, the Sun is located at coordinates $(0,0)$.  }\label{fig8}
   \end{figure*}

Figure~\ref{fig7} shows the projection of the interpolated values on the ($x$,$y$) plane for different $z$ values (i.e. similar to a computerized axial tomography). The thickness of each slice is $\Delta z=10$ pc. Multiple clouds and beams can be observed, which trace small-scale structures. Figure~\ref{fig8} shows vertical slices of the Hydrogen density map for different Galactic longitude values.  The scale indicates density values ranging from $n=0 $ cm$^{-3}$ (red points) to $n=137 $ cm$^{-3}$ (purple points). Note that the four ``corners'' on each plot correspond to regions without X-ray spectra and therefore the interpolation does not provide $n$ values there. An extrapolation scheme, without an estimate of the density profile distribution, can not be performed.  Clouds of different sizes are observed along all l.o.s, an evidence of small-scale structures. In this regime the ISM is found outside equilibrium mainly because the presence of shock fronts associated with Supernovae explosions although the contribution from magnetic fields and cosmic-rays to the physical conditions of the gas are significant \citep{kalb09,kal16}. Observational constraints, such as the densities derived from our analysis, are necessary to compare with the high-resolution 3D hydrodynamical simulations that have been performed in the last decades \citep{avi00,avi01,lag13,gen13}.  

\begin{figure}
\includegraphics[scale=0.36]{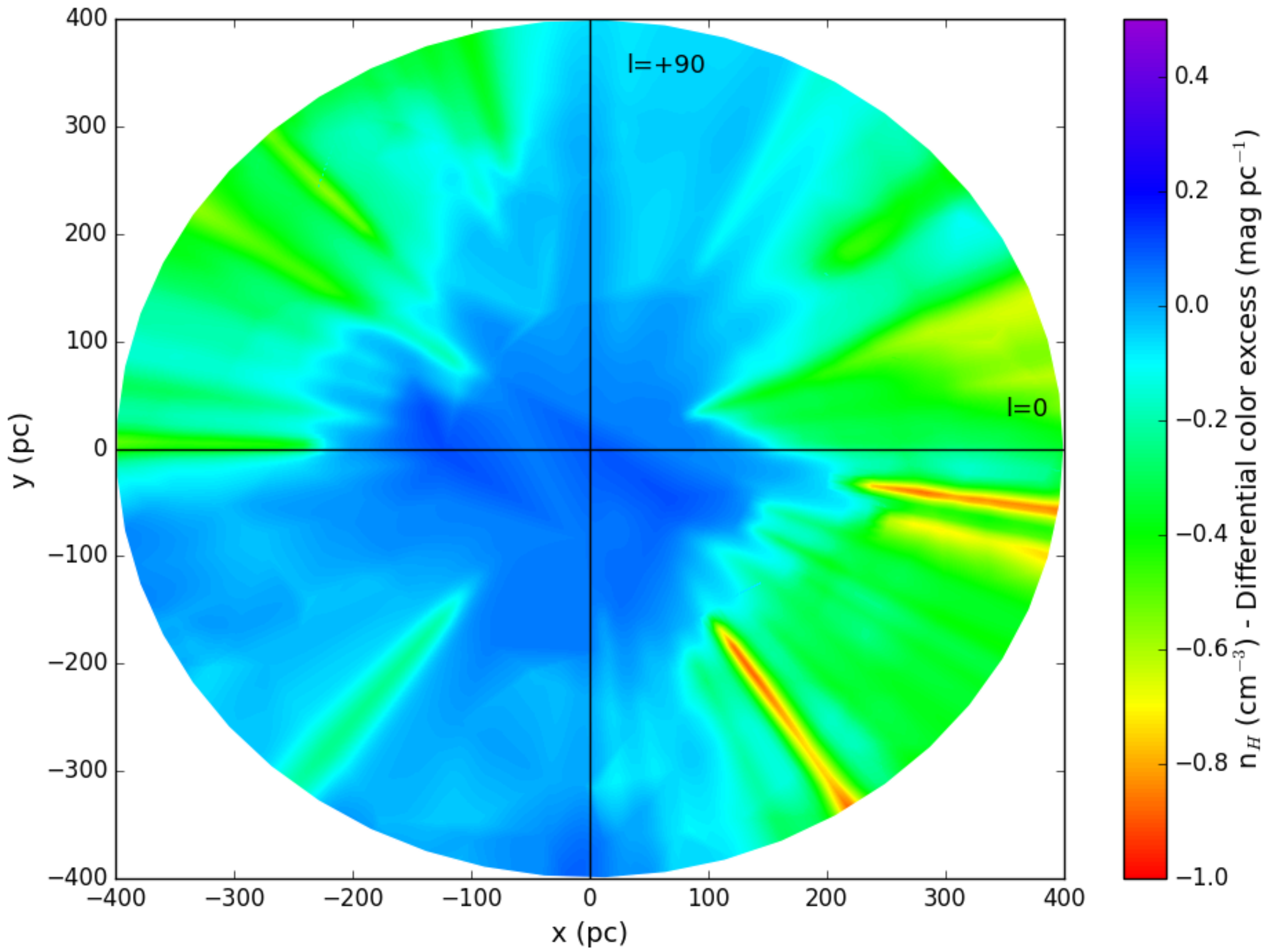} 
      \caption{Subtraction between \citet{lal18} and our results in the Galactic plane. The Sun is located at coordinates $(0,0)$.  }\label{fig9}
   \end{figure}

Figure~\ref{fig9} shows the subtraction between our results and the inverted differential opacity distribution from \citet{lal18} in the Galactic plane. In order to do the subtraction, both results have been normalized to a maximum of $1.0$. The Sun is located at coordinates $(0,0)$.  \citet{lal18} computed their opacity map using $E(B-V)$ measurements in combination with distances obtained from the {\it Gaia} DR1. We found a good agreement between both maps, especially near the center. Larger differences are found in the outer region, where the density errors are larger. It is important to note that the spatial resolution in our maps corresponds to $\sim$ 75~kpc while spatial resolution in \citet{lal18} corresponds to $\sim$ 15~kpc.

\section{Comparison with 21~cm surveys}\label{sec_21cm}
One result from examination of the {\sc EXTraS} catalogue is the comparison between the X-ray derived equivalent column densities and the column densities predicted by the 21~cm maps of the galaxy, as is shown in Figure~\ref{fig4}.  The X-ray columns are larger than the 21~cm columns for a significant fraction  of the sources ($\sim 65\%$). This is in contrast to the behavior expected from to the fact that the X-rays do not traverse the entire galaxy while the 21~cm line should be optically thin for many l.o.s. A similar result has been described by \citet{wil13}, who studied the absorption in the afterglows of gamma ray bursts from the {\it Swift} satellite. There are various possible explanations for this, including: absorption which is intrinsic to the sources or other systematic errors in the X-ray fitting procedure; the existence of absorbing structures in the ISM which are smaller than the angular
resolution of the 21~cm survey beam and which also are rare enough to not skew the net 21~cm emission significantly; and the existence of absorbers which do not manifest themselves as neutral hydrogen but which still affect the X-ray absorption.  We discuss each of these next.

Inherent in the results presented here is the  assumption that the neutral absorption in the {\sc EXTraS} spectra is predominantly intervening between us and the source, so that we can infer the mean density along the l.o.s. If the sources themselves have intrinsic absorption then this assumption is invalid.  Intrinsic absorption is common in classes of X-ray sources which involve compact objects, and is likely associated with outflows or stellar remnants associated with previous evolutionary stages.  The majority of the targets in our sample obtained by correlating with {\it Gaia} are stars, i.e. the X-rays come from an active corona.  X-ray emission from stars is correlated with rotation, and the most X-ray luminous stars are rapidly rotating, and hence relatively young, many in interacting binaries \citep{gud04}.  Spectral observations utilizing both moderate resolution and high resolution instruments shows that intrinsic absorption is not commonly observed from late type stars. Absorption is detected from pre-main sequence stars (FU~Ori and T~Tauri stars) and also from early-type stars in binary systems \citep{gud09}. However, due to the variety of stars that are included in our sample, this effect cannot by itself explain the observed effect.  Other systematic effects which can affect the inferred equivalent column density include incorrect modeling of the underlying continuum; if the true continuum is flatter (i.e. weaker at low energies) than has been assumed then the fitted equivalent column density would be too high in order to suppress the extra continuum. This phenomenon is known from study of extragalactic X-ray spectra, and may indicate a soft component which is not correctly modeled \citep{wil87}.  This possibility is difficult to exclude without careful examination of the goodness-of-fit contours for each source.

Cases where the X-ray equivalent column is greater than the 21~cm column may also be an indication of absorbers other than atomic hydrogen, which can affect the X-ray absorption.  This can include partially ionized gas, i.e. where atomic hydrogen is at least partially ionized but where sufficient {\rm He}~{\sc ii} and partially ionized C, N or O are present to produce X-ray absorption. This would be expected from the warm ionized component (WIM) of the ISM, and is not unlikely given that the surface density of the WIM is estimated to be a significant fraction of that of neutral hydrogen \citep{fer01}.  It is also possible that the greater column could be due to molecular absorption (up to 20\%); molecules in dense or diffuse clouds will have an X-ray cross section which similar to that of atomic gas when viewed with medium resolution instruments.  This too is plausible based on estimates for the surface density of cold molecular gas.  This is the scenario preferred by \citet{wil13}.  Which of these obtains in reality can be determined using high resolution X-ray spectra, which can distinguish the ionization or molecular binding of the absorbers via the strength and energy of the resonance structure near the photoabsorption thresholds of oxygen or other elements. In recent years, multiple analysis of interstellar molecules and dust particles using X-ray spectra have been performed.  These studies include the analysis of X-ray scattering due to intervening interstellar dust grains \citep{lee09,pin13}, detection of absorption features due to molecules such as CO \citep{joa16}, metallic particles \citep{gat16,rog18} and tracers of the dust distribution in the Galactic center \citep{pon15}.

\begin{figure}
\includegraphics[scale=0.35]{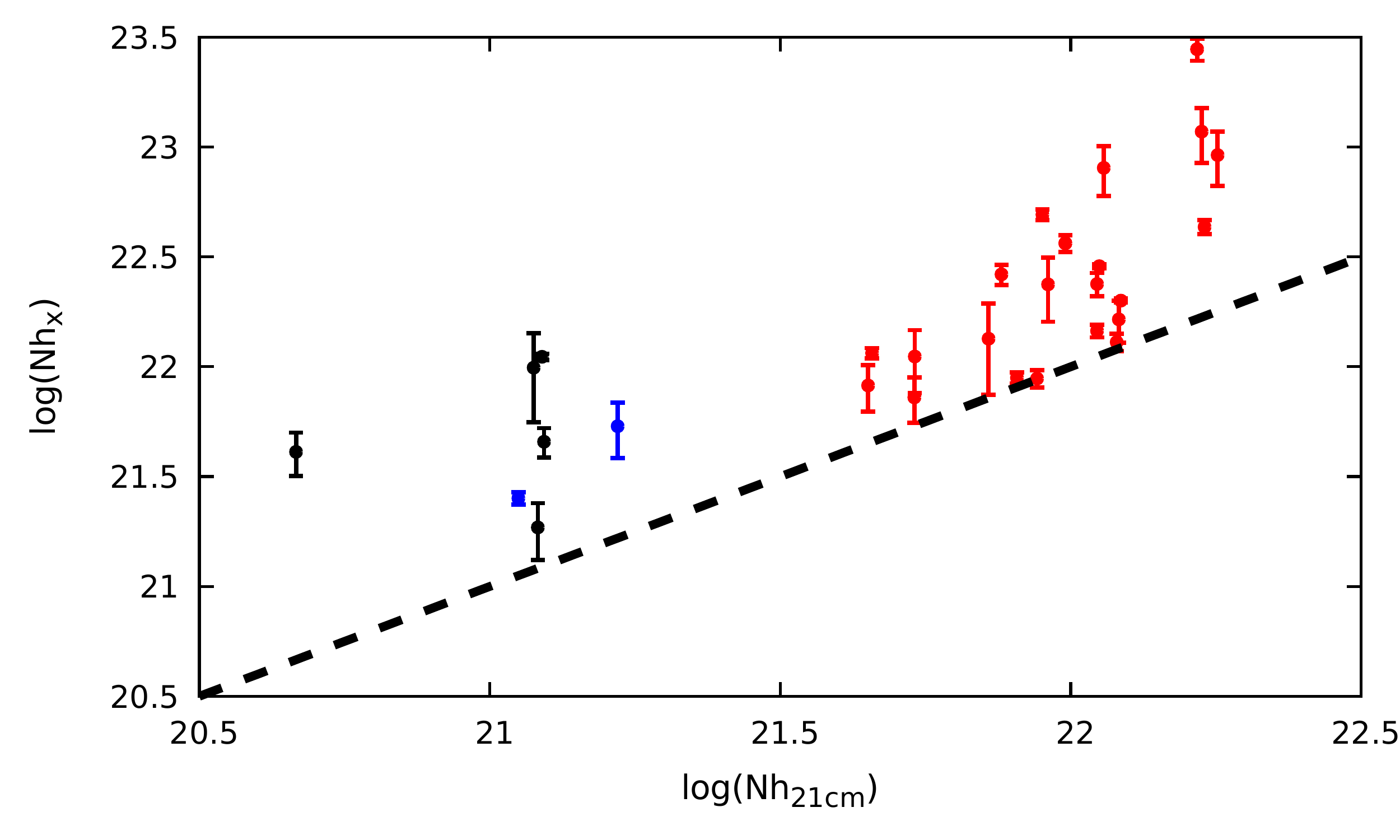}
      \caption{Same as Figure~\ref{fig4} but for lines-of-sight for which molecular clouds have been identified. Colors indicate molecular clouds identified by \citet[black points]{sch14}, \citet[blue points]{hou14} and \citet[red points]{miv17}.  The dashed line indicates the case when $N({\rm H}_{x})$=$N({\rm H}_{21cm})$. }\label{fig_mol_21cm}
   \end{figure}

Figure~\ref{fig_mol_21cm} is similar to Figure~\ref{fig4} but only shows those lines-of-sight for which molecular clouds have been identified and for which $N({\rm H})$ derived from X-ray spectra is larger than 21~cm column densities. Colors indicate molecular clouds identified by \citet[black points]{sch14}, \citet[blue points]{hou14} and \citet[red points]{miv17}. The dashed line indicates the case when $N({\rm H}_{x})$=$N({\rm H}_{21cm})$.  It is expected that, for near molecular clouds, the temperature is too low to be completely traced by 21~cm measurements. There are few surveys available to perform such comparison for low distances.  By using the second {\it Gaia} data release a more complete comparison with multiple molecular surveys covering larger distances will be achievable.    

An additional possible explanation for differences between the X-ray and the 21~cm absorption is due to differing effective beam sizes.  The 21~cm measurements come from a beam size 0.7 deg$^2$ \citep{kal05}, corresponding to a physical size $\simeq$2 pc at a distance of 1~pc. The X-ray absorption effectively probes a region with size limited by the size of the distant source, i.e. 1~AU or so for the largest stellar sources.   A structure of this size would require a very large over-density compared with the typical ISM neutral medium density in order to have a measurable X-ray absorbing column.
The required over-density would be smaller, but still substantial, for structure which is intermediate in size, say 0.01~pc.  Furthermore, such structures would have to be relatively rare, i.e. have a covering fraction relative to our l.o.s $\leq$ 0.1, in order to not skew the 21~cm columns significantly.  The fact that the X-ray and 21~cm columns disagree in a significant fraction of the cases makes this appear implausible, although the fact that the disagreement spans both cases where X-ray column exceeds 21~cm column and vice versa suggests the possibility of such enhancements being common.

It is important to note that the inclusion of the distance measurements from the next {\it Gaia} data releases, in combination with the {\sc EXTraS} catalogue, will allow us to increase the data sample, thus providing an unprecedented detailed mapping of the cold Galactic ISM as observed by X-ray observations including complex structures such as the {\rm H}~{\sc i} Galactic spiral distribution which is difficult to observe using emission line tracers \citep{str07,mcc10}. In this sense, our technique will be better for more distant sources than the stellar color excess technique. Finally, the X-ray all-sky survey to be performed by the {\it eROSITA} instrument on board of the Spectrum Roentgen-Gamma (SRG) satellite \citep{pre06,pre10}, will require accurate Hydrogen column density measurements and distances, as those provided here, in order to estimate X-ray luminosities \citep{pil12}.

\section{Caveats and limitations}\label{sec_cav}
\begin{figure}
\includegraphics[scale=0.32]{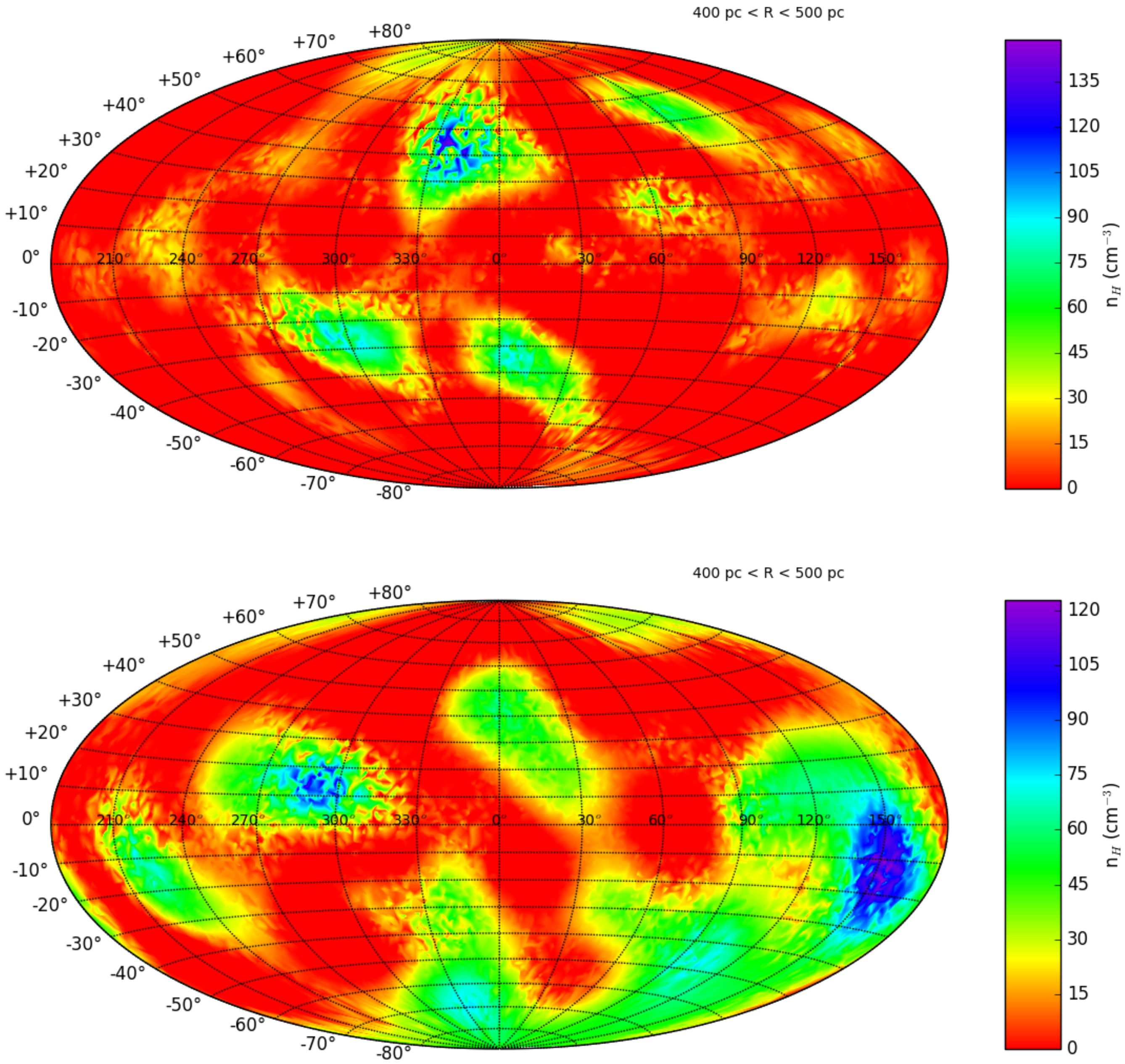}
 
      \caption{  Full-sky 2D map of the density distribution. The map is an Aitoff projection covering distances from $400$ to $500$ pc. Top pannel: density map for a maximum separation of 12.5~arcsecond in the cross-matching between {\sc EXTraS} and {\it Gaia} DR1. Bottom pannel: density map for a maximum separation of 4~arcsecond.   }\label{fig_sys_arcsec}
   \end{figure}

The main caveats in our analysis are as following.
\begin{enumerate}
\item The accuracy of the equivalent column density estimation by X-ray spectral fitting depends on the number of counts. We estimate the expected uncertainties in the column densities for sources with 500 counts in the complete energy range (0.5-10~keV) to be of the order of $\sim 25\% -40\%$, depending on the complexity of the model. For sources with lower number of counts, the uncertainties can be  $>$100$\%$.
\item The nature of the source spectrum will impact the equivalent column density measurements accuracy. For example, a soft-dominated spectrum will give more accurate measurements at low equivalent column density than a harder one. Even when the {\sc EXTraS} catalogue includes a family of continuum models it is plausible to have incorrect modeling for some sources, affecting the equivalent column density measurement obtained.
\item It is know that X-ray absorption is sensitive to species such as {\rm He}~{\sc II} and {\rm H}$_{ 2}$ while 21~cm surveys are not. Such species may explain discrepancies between the X-ray-derived ${\rm H}$ values and the 21~cm or molecular values.
\item Possible variations in the metallic abundances through different l.o.s are not considered in the X-ray fitting procedure. However, large systematic departures from the average values would be needed in order to affect the majority of our measurements.
\item The maximum angular separation in the cross-matching between {\sc EXTraS} and {\it Gaia} DR1 sources affects the final sample and therefore the density map obtained. We use a maximum separation of 12.5~arcsecond, which corresponds to the point-spread function (PSF) of the EPIC PN instrument on board of {\it XMM-Newton}\footnote{\url{https://heasarc.nasa.gov/docs/xmm/uhb/onaxisxraypsf.html}}. Figure~\ref{fig_sys_arcsec} shows an Aitoff projection of the density distribution covering distances from $400$ to $500$ pc (similar to Figure~\ref{fig6}). Top pannel corresponds to a cross-matching with maximum separation of 12.5~arcsecond (i.e. the sample described in Section~\ref{sec_dat}) while the bottom pannel shows the density map for a sample with a maximum separation of 4~arcsecond, a value used in similar cross-matching surveys \citep[see for example ][]{cac08,pin11,lan17}.  We noted that when using 4~arcsecond of separation, the majority of the sources excluded  are located near the Galactic plane, which correspond to a higher column density region. In that case, the sample has column densities $N({\rm H})< 2.3\times 10^{22}$ cm$^{-2}$ while the original sample has column densities up to $ 3.3\times 10^{23}$ cm$^{-2}$. Although lower angular separation corresponds to a more accurate source identification, we decide to use a 12.5~arcsecond separation in the analysis in order to have a better statistic and to cover a large column densities range.

\end{enumerate}

 \section{Summary and Conclusions}\label{sec_con}
The fitting of X-ray spectra provides an accurate way to measure hydrogen equivalent column density values $N({\rm H})$, a very important parameter in multiple astronomy fields.  First, we have analyzed the ${\rm H}$ density distribution in the local ISM using the {\it XMM-Newton} spectral-fit database ({\sc EXTraS}) and the distances estimated by \citet{ast16a} using the first {\it Gaia} data release. {\sc EXTraS} include multiple fits for each sources, allowing an accurate determination of the hydrogen equivalent column densities, which depends on the continuum modeling of the spectra. After the {\it Gaia} - {\it XMM-Newton} cross-correlation, a total of 21946 unique sources have been identified. 

We then used the method explained in \citet{rez17} to infer Hydrogen density distribution from the Hydrogen equivalent column densities, in order to obtain a 3D map of the neutral gas in the ISM. For the final data sample we use those sources that have $<20\%$ of distance uncertainties, $\Delta N({\rm H})<50\%$ and number of counts $>10^{2}$ (2128 sources). In this sense, we are mapping a  local region, with distances lower than 600~pc. Although the projected region is somewhat small, the 3D map shows small-scale density structures. More important, such X-ray density map provides more information about the Hydrogen density distribution compared to 21~cm surveys because X-ray photons trace not only the atomic ISM component but also the ionized gas, molecules and solids.  This is the first time such a map has been created using X-ray spectral fits.   Because systematic uncertainties due to the $N({\rm H})$ dependence on the continuum fitting model and the source identifications, the present maps should be considered qualitatively at this point. A similar analysis is currently under development using the much more complete and accurate {\it Gaia} DR2, which we will be feature in a future publication.

\section*{Acknowledgements}
EG acknowledge support by the DFG cluster of excellence `Origin and Structure of the Universe'. AK, MO, and JW acknowledge partial funding from the European Union's Seventh Framework Programme under grant agreement number 607452. We thank Rosine Lallement for providing the opacity distribution data in the Galactic plane.

\bibliographystyle{mnras}


\end{document}